\documentclass[twocolumn,pra,showpacs,preprintnumbers]{revtex4-1}%

\usepackage{booktabs}
\usepackage{amsmath, amssymb, amsfonts}
\usepackage{bm}
\usepackage{dsfont}
\usepackage{soul}
\usepackage{graphicx}
\usepackage{epsfig}
\usepackage{subfigure}
\usepackage{multirow}
\usepackage{makecell}
\usepackage[english]{babel}
\usepackage{mathrsfs}
\usepackage{lipsum}
\usepackage{appendix}
\usepackage{courier}
\usepackage[colorlinks,linkcolor=blue,citecolor=blue]{hyperref}%
\providecommand{\U}[1]{\protect \rule{.1in}{.1in}}
\begin{document}

\title{Configuration-based understanding of superradiant phase transitions in Dicke lattices}
\author{Peng-Fei Wei$^{1}$}
\author{Zhihai Wang$^{1}$}
\email{wangzh761@nenu.edu.cn}
\affiliation{1. Center for Quantum Sciences and School of Physics, Northeast Normal University, Changchun 130024, China}

\begin{abstract}

The emergence of multiple superradiant phases in Dicke lattice models has attracted considerable attention in the quantum optics community. However, a unified understanding of the origin of multistability and its relation to different superradiant phases is still lacking. Here, we develop a configuration-based understanding to classify the superradiant phases in Dicke lattices. We show that photon hopping naturally organizes the possible superradiant configurations according to the lattice symmetry, providing a unified interpretation of the nonequilibrium phase diagram and the emergence of multistability. For the dissipative four-site Dicke lattice, we obtain the complete phase diagram and identify the coexistence of up to four stable superradiant phases. The proposed classification is further extended to five- and six-site lattices. Moreover, we demonstrate that the same configuration-based understanding also applies to the closed Dicke lattice, where the ground state uniquely selects one of the allowed configurations. Finally, we show that different configurations may belong to either same or distinct nonequilibrium universality classes in the dissipative Dicke lattice, while they share the same equilibrium universality class in the closed Dicke lattice. Our results provide a unified picture for understanding equilibrium and nonequilibrium superradiant phase transitions in Dicke lattices.
\end{abstract}

\maketitle

\section{Introduction}

The dissipative quantum phase transition (QPT) in open quantum systems has attracted tremendous attention over the past two decades~\cite{Werner2005,Nesterov2008,Morrison2008,OS2010,OS2012b,Carmichael2015,Fitzpatrick2017,Hwang2018,Minganti2018,
Carollo2020,Rossini2021,Bibak2023,Belyansky2025,Ali2026}, extending the concept of quantum criticality from condensed-matter physics to quantum optical platforms. A paradigmatic example is the open Dicke model~\cite{Dimer2007,Nagy2011,DallaTorre2013,DallaTorre2016,Dicke2017,Gelhausen2017,
Boneberg2022,Muller2025}, in which an ensemble of two-level atoms collectively couples to a lossy resonator mode. When the atom-resonator coupling exceeds a critical value, the system undergoes a transition from the normal phase (NP), where both the atoms and the resonator  remain in their ground states, to the superradiant phase (SRP), characterized by macroscopic occupations of both atomic and photonic excitations. Along with extensive theoretical studies of the Dicke model~\cite{Dicke1954,Dicke19731,Dicke19732,Dicke19733,Dicke2003PRL,Dicke2003PRE,Emary2004,Dicke2010,Dicke2014,Zhang2018,
Dicke2019,Dicke2020,Dicke2023,Puel2024,Han2024,Mendonca2025}, both equilibrium and dissipative superradiant phase transitions have been experimentally observed on a variety of physical platforms~\cite{DickeEX2010,DickeEX2014,Hamner2014,DickeEX2015,DickeEX2021,DickeEX2023}.

An intriguing feature of the open Dicke model is the emergence of multistability~\cite{Keeling2010,Dicke2012,Dicke2018,Dicke2024,Zhu2024}, where multiple stable steady states coexist under the same system parameters and the final state depends on the initial condition. In the single-cavity Dicke model, multistability originates from nonlinear interactions induced by Bose--Einstein condensates, leading to the coexistence of the NP and the SRP~\cite{Keeling2010,Dicke2012}. Introducing photon hopping between neighboring cavities further enriches the nonequilibrium phase structure in Dicke lattices. For a two-site lattice, the coexistence of homogeneous and antisymmetric SRPs gives rise to bistability~\cite{twoD2024,twoD2025}. For three-site lattices, periodic boundary conditions support the coexistence of homogeneous and inhomogeneous SRPs, whereas open boundary conditions further allow tristability among different inhomogeneous phases~\cite{NDickeWei}. By contrast, the closed Dicke lattice possesses a unique ground-state SRP determined by energy minimization. For odd-site lattices, this leads to the emergence of frustrated SRP~\cite{threeD2022,threeD2023,threeD2025}. Despite these remarkable discoveries, the existing studies have largely focused on specific lattice sizes and geometries, and a unified understanding of the origin of different SRPs and their relationship remains absent.

In this work, we develop a configuration-based framework for understanding SRP in Dicke lattices. We show that, once photon hopping is introduced, all possible superradiant configurations can be classified according to the lattice symmetry. This classification provides a unified interpretation of the different SRPs and naturally explains the emergence of multistability as the simultaneous stabilization of multiple configuration classes. In the closed Dicke lattice, by contrast, the ground state uniquely selects one of the allowed configurations through energy minimization. Based on this framework, we establish a direct connection between the spatial configurations of the SRP, the equilibrium and nonequilibrium phase diagrams.

Specifically, we first classify all possible superradiant configurations for finite Dicke lattices and obtain the complete phase diagram of the four-site lattice under periodic boundary conditions. We identify the emergence of fourfold multistability in the dissipative system and further generalize the classification to larger lattices with five and six sites. We then investigate the closed Dicke lattice, where the ground state is shown to select either the ferromagnetic or antiferromagnetic configuration depending on the sign of the photon hopping. Finally, by comparing the equilibrium and nonequilibrium phase transitions, we demonstrate that different superradiant configurations can correspond to same or distinct universality classes in the dissipative Dicke lattice, while they belong to the same universality class in the closed Dicke lattice.

The remainder of this paper is organized as follows. In Sec.~\ref{model}, we introduce the Dicke lattice model together with the configuration classification. Section~\ref{phaseD} presents the nonequilibrium phase diagram and multistability of the dissipative four-site Dicke lattice. The equilibrium phase diagram and the corresponding superradiant phase transitions of the closed Dicke lattice are discussed in Sec.~\ref{closed}. Finally, we summarize our results in Sec.~\ref{Con}. Technical details of the steady-state and ground-state analyses are presented in Appendices~\ref{Aopen} and \ref{Bclosed}, respectively.

\begin{table*}[htbp]
\centering
\caption{Classification of superradiant configuration classes under lattice symmetries.}
\renewcommand{\arraystretch}{2} 
\setlength{\tabcolsep}{12pt}
\begin{tabular}{l lll}
\toprule
Sites($N$)

& \makecell[l]{Representative configurations and class degeneracies $(m)$}&
\makecell[l]{Number of\\ classes ($n$)}&
\makecell[l]{Count\\($\sum_{i=1}^{n}m_{i}$)} \\
\midrule

$N=3$
& \texttt{\textbf{[+\,+\,+]}}~($m_{1}=2$),~\texttt{\textbf{[+\,-\,-]}}~($m_{2}=6$) & $n=2$ & $2^3$ \\
\addlinespace[1.2ex]
$N=4$
& \makecell[l]{\texttt{\textbf{[+\,+\,+\,+]}}~($m_{1}=2$),~\texttt{\textbf{[+\,-\,-\,-]}}~($m_{2}=8$)\\
\texttt{\textbf{[+\,+\,-\,-]}}~($m_{3}=4$),~\texttt{\textbf{[+\,-\,+\,-]}}~($m_{4}=2$)} & $n=4$ & $2^4$ \\
\addlinespace[1.2ex]
$N=5$
& \makecell[l]{\texttt{\textbf{[+\,+\,+\,+\,+]}}~($m_1=2$),~\texttt{\textbf{[+\,-\,-\,-\,-]}}~($m_2=10$)\\
\texttt{\textbf{[+\,+\,-\,-\,-]}}~($m_3=10$),~\texttt{\textbf{[+\,-\,+\,-\,-]}}~($m_4=10$)} & $n=4$ & $2^5$ \\
\addlinespace[1.2ex]
$N=6$
& \makecell[l]{\texttt{\textbf{[+\,+\,+\,+\,+\,+]}}~($m_1=2$),~\texttt{\textbf{[+\,-\,-\,-\,-\,-]}}~($m_2=12$),~
\texttt{\textbf{[+\,+\,-\,-\,-\,-]}}~($m_3=12$)\\
\texttt{\textbf{[+\,-\,+\,-\,-\,-]}}~($m_4=12$),~
\texttt{\textbf{[+\,-\,-\,+\,-\,-]}}~($m_5=6$),~\texttt{\textbf{[+\,+\,+\,-\,-\,-]}}~($m_6=6$)\\
\texttt{\textbf{[+\,+\,-\,+\,-\,-]}}~($m_7=12$),~\texttt{\textbf{[+\,-\,+\,-\,+\,-]}}~($m_8=2$)}
 & $n=8$ & $2^6$\\
\bottomrule
\end{tabular}
\label{tab:N}
\end{table*}

\section{Dissipative Dicke lattice model}\label{model}

We consider a one-dimensional dissipative Dicke lattice composed of $N$ coupled atom-resonator units. Each lattice site consists of a single-mode resonator collectively coupled to an ensemble of $N_a$ identical two-level atoms, while neighboring resonators are connected through coherent photon hopping. The total Hamiltonian is given by
\begin{equation}
H=
\sum_{i=1}^{N} H_i^{\rm Dicke}
-
\xi
\sum_{i=1}^{N}
\left(
c_i^\dagger c_{i+1}
+
c_{i+1}^\dagger c_i
\right),
\label{eq:H}
\end{equation}
where periodic boundary conditions are imposed through $c_{N+1}=c_1$. The local Hamiltonian at site $i$ takes the form
\begin{equation}
H_i^{\rm Dicke}
=
\omega_c c_i^\dagger c_i
+
\omega_a S_i^z
+
\frac{2g}{\sqrt{N_a}}
\left(
c_i+c_i^\dagger
\right)
S_i^x.
\end{equation}

Here, $c_i$ ($c_i^\dagger$) is the annihilation (creation) operator of the resonator mode with resonance frequency $\omega_c$, while
$S_i^{(x,y,z)}=\sum_{j=1}^{N_a}\sigma_{i,j}^{(x,y,z)}/2$ denotes the collective spin operator of the atomic ensemble with transition frequency $\omega_a$. The parameter $g$ characterizes the collective atom-resonator coupling strength, and $\xi$ describes coherent photon hopping between adjacent resonators.

To account for the inevitable photon dissipation in cavity QED platforms, we consider resonator losses as the dominant decoherence channel. The nonequilibrium dynamics of the system is therefore governed by the Lindblad master equation
\begin{equation}
\frac{d\rho}{dt}
=
-i[H,\rho]
+
\kappa
\sum_{i=1}^{N}
\left(
2c_i\rho c_i^\dagger
-
c_i^\dagger c_i\rho
-
\rho c_i^\dagger c_i
\right),
\label{eq:mastereq}
\end{equation}
where $\kappa$ denotes the photon decay rate of each resonator.

Dissipative phase transitions in Dicke lattice models have been extensively studied in Refs.~\cite{twoD2024,twoD2025,NDickeWei}. In contrast to their equilibrium counterparts, these transitions are encoded in the nonequilibrium steady states of the dissipative dynamics and often give rise to SRP with multiple stable configurations. Before turning to the quantitative analysis of our model, we first clarify the physical mechanism behind this multistability by classifying the possible steady-state configurations of the lattice.

As a starting point, we first consider the decoupled limit with $\xi=0$, where each lattice site behaves as an independent dissipative Dicke model. In this limit, every site possesses its own local $\mathbb{Z}_2$ symmetry, which is spontaneously broken once the atom-resonator coupling exceeds the critical value
\begin{equation}
g_c=\frac{1}{2}\sqrt{\omega_a\omega_c\left(1+\frac{\kappa^2}{\omega_c^2}\right)}.
\end{equation}
Above this threshold, each resonator develops a finite coherent field with
${\rm Re}\langle c_i\rangle$ taking either a positive or a negative value, corresponding to two symmetry-related superradiant branches. Consequently, an array of $N$ uncoupled sites possesses $2^N$ degenerate superradiant configurations, each specified by the sign distribution of the local order parameters.

The situation changes qualitatively once coherent photon hopping ($\xi\neq0$) is introduced. The independent local $\mathbb{Z}_2$ symmetries are reduced to a global $\mathbb{Z}_2$ symmetry, while the ring geometry is additionally invariant under cyclic lattice translations,
\begin{equation}
1\rightarrow2\rightarrow3\rightarrow\cdots\rightarrow N\rightarrow1.
\end{equation}
The dissipative superradiant phase transition considered below is accompanied by the spontaneous breaking of these symmetries.

Because of these symmetry constraints, the original $2^N$ configurations are no longer all equivalent. Instead, they can be grouped into distinct configuration classes, whereas the configurations within the same class remain degenerate. The complete classification for $N=3$--$6$ is summarized in Table~\ref{tab:N}. Each configuration class represents a distinct pattern of local superradiant order and therefore provides a candidate steady-state configuration of the lattice. Depending on the system parameters, several of these classes can become dynamically stable simultaneously, giving rise to multistability. This picture naturally explains the bistability previously reported in the Dicke dimer ($N=2$)~\cite{twoD2024} and trimer ($N=3$)~\cite{NDickeWei}. As we will show below, increasing the lattice size further enriches the nonequilibrium phase diagram, leading to tristability and four-fold stability for $N=4$, with a straightforward generalization to larger lattices such as $N=5$ and $N=6$. In this way, the seemingly complicated multistability can be understood from a unified symmetry-based classification of the underlying superradiant configurations.

\section{Dissipative Superradiant Phase Transition}\label{phaseD}

In this section, we determine the nonequilibrium steady-state phase diagram of the dissipative Dicke lattice in the thermodynamic limit, $N_a\rightarrow\infty$, using a mean-field approach. The equation of motion for the expectation value of an arbitrary operator $\mathcal{O}$ follows from
$d\langle \mathcal{O}\rangle/dt={\rm Tr}\left(\mathcal{O}\dot{\rho}\right)$.
Within the mean-field approximation, operator correlations are factorized as
$\langle AB\rangle\simeq \langle A\rangle\langle B\rangle$. We introduce the rescaled mean-field variables
$\langle c_j\rangle=\sqrt{N_a}\left(R_j+iI_j\right)$,
$\langle S_j^{-}\rangle=N_a\left(R_j^{s}+iI_j^{s}\right)$, and
$\langle S_j^{z}\rangle=N_a S_j^{z}$. The resulting semiclassical equations of motion are
\begin{equation}
\begin{aligned}
\dot R_j &=
-\kappa R_j+\omega_c I_j-\xi\left(I_{j+1}+I_{j-1}\right),\\
\dot I_j &=
-\omega_c R_j-\kappa I_j-2g R_j^{s}
+\xi\left(R_{j+1}+R_{j-1}\right),\\
\dot R_j^{s} &=
\omega_a I_j^{s},\\
\dot I_j^{s} &=
-\omega_a R_j^{s}+4g R_j S_j^{z},\\
\dot S_j^{z} &=
-4g R_j I_j^{s}.
\end{aligned}
\label{eq:ss}
\end{equation}

The steady-state solutions
$\{R_j^{\rm ss},I_j^{\rm ss},(R_j^s)^{\rm ss},(I_j^s)^{\rm ss},(S_j^z)^{\rm ss}\}$
are obtained by setting the left-hand side of Eq.~\eqref{eq:ss} to zero, together with the spin-length constraint
$|\langle S_j^-\rangle|^2+\langle S_j^z\rangle^2=N_a^2/4$. To identify the dynamically stable steady states, we linearize the mean-field equations around each solution by writing
\begin{equation}
A_j=A_j^{\rm ss}+\delta A_j,
\qquad
A_j\in\{R_j,I_j,R_j^s,I_j^s,S_j^z\}.
\end{equation}
Keeping only terms linear in the fluctuations yields
\begin{equation}
\frac{d}{dt}\delta\boldsymbol{\Psi}
=
M\,\delta\boldsymbol{\Psi},
\end{equation}
where $\delta\boldsymbol{\Psi}$ collects all fluctuation variables and $M$ is the corresponding coefficient matrix, whose explicit form is given in Appendix~\ref{Aopen}. A steady-state solution is dynamically stable if all eigenvalues of $M$ have negative real parts, in accordance with the Routh--Hurwitz criterion~\cite{QZZL1987}. Therefore, the nonequilibrium phase diagram is obtained by identifying all steady-state solutions of Eq.~\eqref{eq:ss} and retaining only those satisfying the stability condition. In what follows, we mainly present the phase diagram and universality classes for the four-site Dicke lattice ($N=4$), and then briefly discuss the steady-state configurations for $N=5$ and $N=6$.

\subsection{Analytical solutions for $N=4$}

A complete closed-form solution of Eq.~\eqref{eq:ss} is generally unavailable because the steady-state equations are nonlinear. Nevertheless, analytical solutions can be derived for several representative configurations, whose stability boundaries can be expressed in terms of the atom--resonator coupling strength.

The simplest steady-state solution is the trivial one,
\begin{equation}
R_j=I_j=R_j^s=I_j^s=0,
\qquad
S_j^z=-\frac{1}{2},
\end{equation}
which exists for every site $j$. This solution corresponds to the NP, where the photonic fields vanish and all atomic ensembles remain in their ground state. Linear stability analysis shows that the NP is stable for $g<g_c^{\rm NP}$, with the critical coupling
\begin{equation}
g_c^{\rm NP}
=
\min_k
\frac{1}{2}
\sqrt{
\omega_a \omega_k
\left(
1+\frac{\kappa^2}{\omega_k^2}
\right)
}.
\label{eq:NPB}
\end{equation}
Here, $\omega_k$ denotes the normal-mode frequency obtained by diagonalizing the photonic hopping Hamiltonian, given by
\begin{equation}
\omega_k=\omega_c-2\xi\cos\left[\frac{\pi(k-1)}{2}\right],
\qquad k=1,2,3,4 .
\end{equation}
Requiring all photonic normal modes to have positive frequencies, $\omega_k>0$, imposes the condition $-\omega_c/2<\xi<\omega_c/2$.

For stronger atom-resonator coupling $g>g_c^{\rm NP}$, nontrivial steady states with finite photonic fields and finite transverse spin coherence can become stable. These solutions correspond to SRPs. Different SRPs are distinguished by their spatial configurations, which can be characterized by the local order parameters, namely the real photonic components $R_j$. In the following, we present the analytical expressions for representative configurations, while leaving the detailed derivations to Appendix~\ref{Aopen}.

\textit{(1) Homogeneous superradiant phase} (HSRP).
The HSRP is characterized by a spatially uniform order parameter,
$R_1^{\rm ss}=R_2^{\rm ss}=R_3^{\rm ss}=R_4^{\rm ss}
\equiv \mathcal{R}_1$, with the remaining mean-field variables also being site independent. This phase corresponds to the configuration $\texttt{\textbf{[+\,+\,+\,+]}}$ in Table~\ref{tab:N} and is analogous to a ferromagnetic ordering pattern. Owing to the spontaneous breaking of the global $\mathbb{Z}_2$ parity symmetry, the corresponding steady state is twofold degenerate. The HSRP steady-state solution can be written as
\begin{equation}
\begin{aligned}
\mathcal{R}_1
&=
\frac{\omega_a}{4g}
\frac{\mathcal{R}_1^s}{\mathcal{Z}_1},
\,
\mathcal{I}_1
=
\frac{\kappa}{\omega_c-2\xi}\mathcal{R}_1,
\\
\mathcal{R}_1^s
&=
\pm
\mathcal{Z}_1
\sqrt{
\frac{1}{(2\mathcal{Z}_1)^2}-1
},
\,
\mathcal{I}_1^s
=0,
\\
\mathcal{Z}_1
&=
-\frac{\omega_a(\omega_c-2\xi)}{8g^2}
\left[
1+\frac{\kappa^2}{(\omega_c-2\xi)^2}
\right].
\end{aligned}
\label{eq:HSRP}
\end{equation}
As shown in Appendix~\ref{Aopen}, linear stability analysis yields the critical coupling
\begin{equation}
g_c^{\rm HSRP}
=
\max_k
\frac{1}{2}
\left[
\frac{
\omega_a^2\omega_k
\left(\kappa^2+\omega_1^2\right)^3
}{
\omega_1^3
\left(\kappa^2+\omega_k^2\right)
}
\right]^{1/4}.
\end{equation}
The HSRP becomes dynamically stable for $g>g_c^{\rm HSRP}$.

\textit{(2) Inhomogeneous superradiant phase 1} (ISRP1).
The ISRP1 is characterized by an alternating pattern of the real part of the order parameter,
$R_1^{\rm ss}=-R_2^{\rm ss}=R_3^{\rm ss}=-R_4^{\rm ss}\equiv \mathcal{R}_2$.
This phase corresponds to the configuration $\texttt{\textbf{[+\,-\,+\,-]}}$ in Table~\ref{tab:N}, with a twofold-degenerate steady state, and is analogous to an antiferromagnetic ordering pattern. The ISRP1 steady-state solution takes the form
\begin{equation}
\begin{aligned}
\mathcal{R}_2
&=
\frac{\omega_a}{4g}
\frac{\mathcal{R}_2^s}{\mathcal{Z}_2},
\,
\mathcal{I}_2
=
\frac{\kappa}{\omega_c+2\xi}\mathcal{R}_2,
\\
\mathcal{R}_2^s
&=
\pm
\mathcal{Z}_2
\sqrt{
\frac{1}{(2\mathcal{Z}_2)^2}-1
},
\,
\mathcal{I}_2^s
=0,
\\
\mathcal{Z}_2
&=
-\frac{\omega_a(\omega_c+2\xi)}{8g^2}
\left[
1+\frac{\kappa^2}{(\omega_c+2\xi)^2}
\right].
\end{aligned}
\label{eq:ISRP1}
\end{equation}
The ISRP1 becomes dynamically stable for $g>g_c^{\rm ISRP1}$, where
\begin{equation}
g_c^{\rm ISRP1}
=
\max_k
\frac{1}{2}
\left[
\frac{
\omega_a^2\omega_k
\left(\kappa^2+\omega_3^2\right)^3
}{
\omega_3^3
\left(\kappa^2+\omega_k^2\right)
}
\right]^{1/4}.
\end{equation}

\textit{(3) Inhomogeneous superradiant phase 2} (ISRP2).
The ISRP2 corresponds to another symmetry-broken configuration, in which two adjacent sites share the same sign of the real order parameter, while the remaining two adjacent sites have the opposite sign. A representative configuration is $\texttt{\textbf{[+\,+\,-\,-]}}$, with
$R_1^{\rm ss}=R_2^{\rm ss}=-R_3^{\rm ss}=-R_4^{\rm ss}\equiv \mathcal{R}_3$.
This configuration is fourfold degenerate. The ISRP2 steady-state solution is
\begin{equation}
\begin{aligned}
\mathcal{R}_3
&=
\frac{\omega_a}{4g}
\frac{\mathcal{R}_3^s}{\mathcal{Z}_3},
\,
\mathcal{I}_3
=
\frac{\kappa}{\omega_c}\mathcal{R}_3,
\\
\mathcal{R}_3^s
&=
\pm
\mathcal{Z}_3
\sqrt{
\frac{1}{(2\mathcal{Z}_3)^2}-1
},
\,
\mathcal{I}_3^s
=0,
\\
\mathcal{Z}_3
&=
-\frac{\omega_a\omega_c}{8g^2}
\left(
1+\frac{\kappa^2}{\omega_c^2}
\right).
\end{aligned}
\label{eq:ISRP2}
\end{equation}
Interestingly, the steady-state amplitudes in Eq.~\eqref{eq:ISRP2} are independent of the hopping strength $\xi$. The stability boundary, however, still depends on $\xi$ through the photonic normal-mode frequencies $\omega_k$. The ISRP2 becomes dynamically stable for
$g>g_c^{\rm ISRP2}$, where
\begin{equation}
g_c^{\rm ISRP2}
=
\max_k
\frac{1}{2}
\left[
\frac{
\omega_a^2\omega_k
\left(\kappa^2+\omega_c^2\right)^3
}{
\omega_c^3
\left(\kappa^2+\omega_k^2\right)
}
\right]^{1/4}.
\end{equation}

In addition to the analytical solutions discussed above, the classification in Table~\ref{tab:N} suggests another possible configuration, namely $\texttt{\textbf{[+\,-\,-\,-]}}$, which is eightfold degenerate. This configuration, however, does not admit a simple closed-form expression. We therefore identify the corresponding inhomogeneous SRP numerically. This phase is denoted as ISRP3, and its properties will be characterized in the following.

\subsection{Steady-state phase diagram}

\begin{figure}
    \centering
    \includegraphics[width=\linewidth]{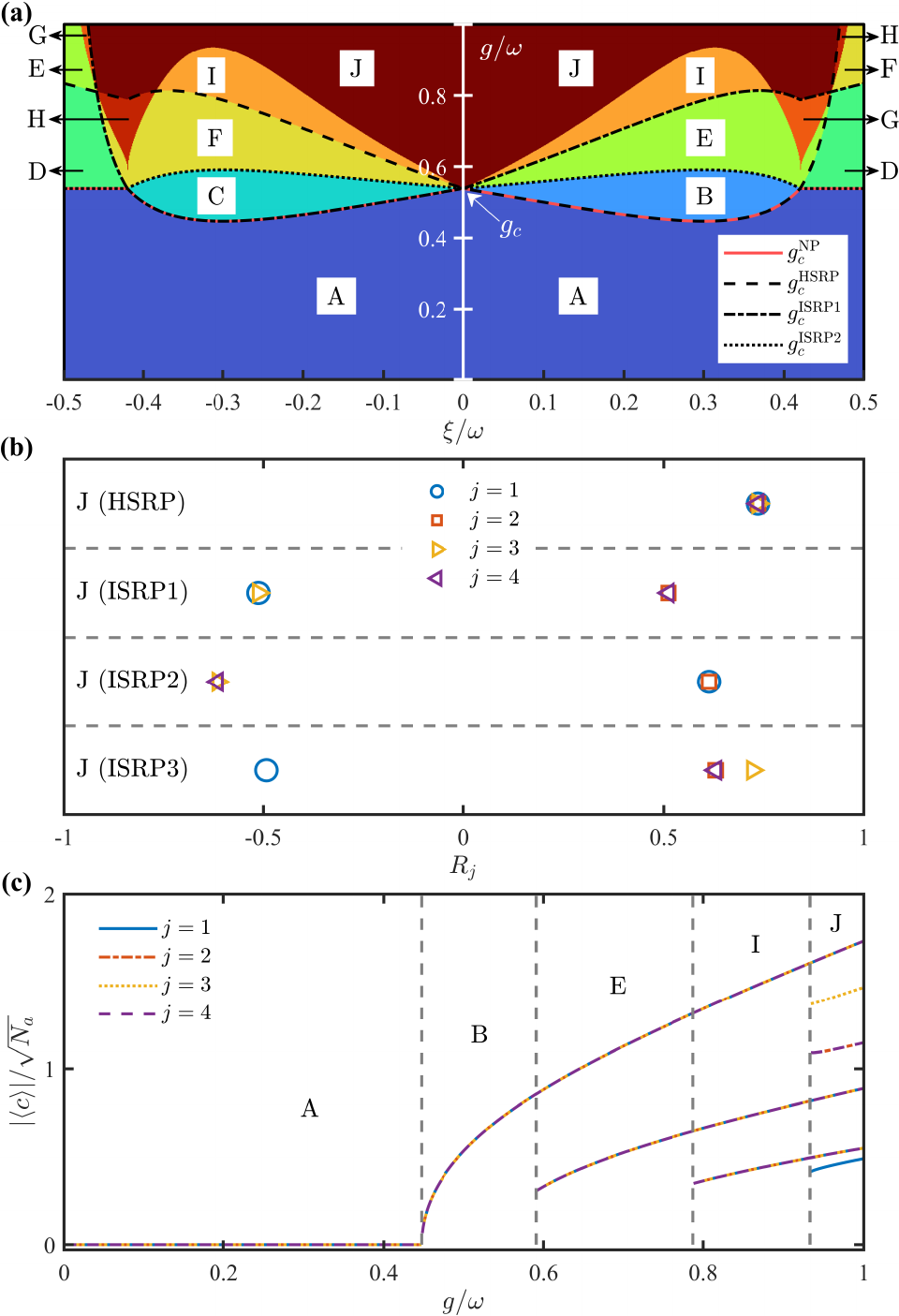}
    \caption{(a) Phase diagram for the open Dicke lattice model with $N=4$. (b) The real part of order parameter $R_j$ about each site with $\xi=0.1\omega$ and $g=0.8\omega$. (c) Plot of the order parameter $|\langle c_{j}\rangle|/\sqrt{N_a}=\sqrt{R_J^2+I_J^2}$ as a function of the coupling strength $g/\omega$ with $\xi=0.3\omega$. Other parameters are set to $\omega_a=\omega_c=\omega$ and $\kappa=0.4\omega$.}
    \label{fig:open}
\end{figure}

Starting from Eq.~\eqref{eq:ss}, we numerically solve the nonlinear steady-state equations and determine the stability of each solution through linear stability analysis. The resulting nonequilibrium phase diagram is presented in Fig.~\ref{fig:open}(a), while the corresponding phase characteristics are summarized in Table~\ref{tab:phase}. The NP, labeled A, is separated from the SRP by the critical boundary $g_c^{\rm NP}$, indicated by the solid curves in Fig.~\ref{fig:open}(a).

The analytical solutions obtained in the previous subsection, namely the HSRP, ISRP1, and ISRP2, are fully confirmed by the numerical calculations together with their corresponding stability boundaries. Throughout this work, symmetry-related degenerate steady states are regarded as belonging to the same phase. Consequently, the analytically derived phases appear not only as the monostable regions B, C, and D, but also as the building blocks of the multistable regions E--J. As summarized in Table~\ref{tab:phase}, regions E and F correspond to the coexistence of two SRPs, regions G, H, and I contain three stable SRPs, whereas region J hosts all four SRPs simultaneously, representing the maximal multistability for the four-site Dicke lattice. This classification provides a unified interpretation of the nonequilibrium phase diagram.

Another prominent feature of Fig.~\ref{fig:open}(a) is its asymmetry with respect to the sign of the photon hopping strength $\xi$. For $\xi>0$, the HSRP occupies a substantially larger region than the ISRP1, whereas the opposite behavior is found for $\xi<0$. This asymmetry can be understood from the analytical expressions of the critical couplings: the HSRP is governed by the lowest photonic normal mode $\omega_1=\omega_c-2\xi$, while the ISRP1 is determined by the highest normal mode $\omega_3=\omega_c+2\xi$. Consequently, positive (negative) hopping lowers the threshold of the HSRP (ISRP1), thereby enlarging its stability region.

In the multistable regions G, H, and J, we further identify the ISRP3 numerically. Figure~\ref{fig:open}(b) presents a representative parameter point in region J, where the HSRP, ISRP1, ISRP2, and ISRP3 coexist simultaneously, demonstrating the emergence of fourfold multistability. Unlike the HSRP, ISRP1, and ISRP2, the ISRP3 exhibits a lower-symmetry spatial structure, in which only one pair of lattice sites possesses identical photon amplitudes, while the remaining two sites are inequivalent. For the representative configuration shown in Fig.~\ref{fig:open}(b), the order parameters satisfy
$R_1<0$, $R_2>0$, $R_3>0$, $R_4>0$, together with
$R_2=R_4$. The numerical results for the HSRP, ISRP1, and ISRP2 agree excellently with the corresponding analytical solutions, thereby confirming the validity of the analytical stability analysis.

To further illustrate how the multistable phases emerge, Fig.~\ref{fig:open}(c) shows a representative cut of the phase diagram at $\xi=0.3\omega$. As the atom--resonator coupling strength increases, additional superradiant configurations become dynamically stable, and the system successively undergoes the phase sequence
${\rm A}\rightarrow{\rm B}\rightarrow{\rm E}\rightarrow{\rm I}\rightarrow{\rm J}$.
Correspondingly, the number of coexisting superradiant phases increases from one to two, three, and finally four, in agreement with the phase characterization summarized in Table~\ref{tab:phase}. While the transition from the NP to a SRP remains continuous, the transitions between different SRPs are governed by the competition among multiple stable steady-state branches and may therefore be either continuous (cross over) or discontinuous (phase transition).

\begin{table}[t]
\centering

\caption{The character of phase with Dicke lattice model for $N=4$.}
\renewcommand{\arraystretch}{1.2}
\setlength{\tabcolsep}{5.7pt}
\begin{tabular}{l ccccccccc}
\toprule

& B & C & D & E & F & G & H & I & J \\
\midrule
\makecell[l]{number of\\steady states}
& 1 & 1 & 1 & 2 & 2 & 3 & 3 & 3 & 4\\
HSRP
& $\checkmark$ &      &     & $\checkmark$  &      & $\checkmark$ &     & $\checkmark$ & $\checkmark$ \\
ISRP1
&      & $\checkmark$ &     &       & $\checkmark$ &      & $\checkmark$ & $\checkmark$ & $\checkmark$ \\
ISRP2
&      &     & $\checkmark$ & $\checkmark$  & $\checkmark$ & $\checkmark$ & $\checkmark$ & $\checkmark$ & $\checkmark$ \\
ISRP3
&      &      &      &       &      & $\checkmark$ & $\checkmark$ &      & $\checkmark$ \\
\bottomrule
\end{tabular}
\label{tab:phase}
\end{table}

\begin{figure}
    \centering
    \includegraphics[width=\linewidth]{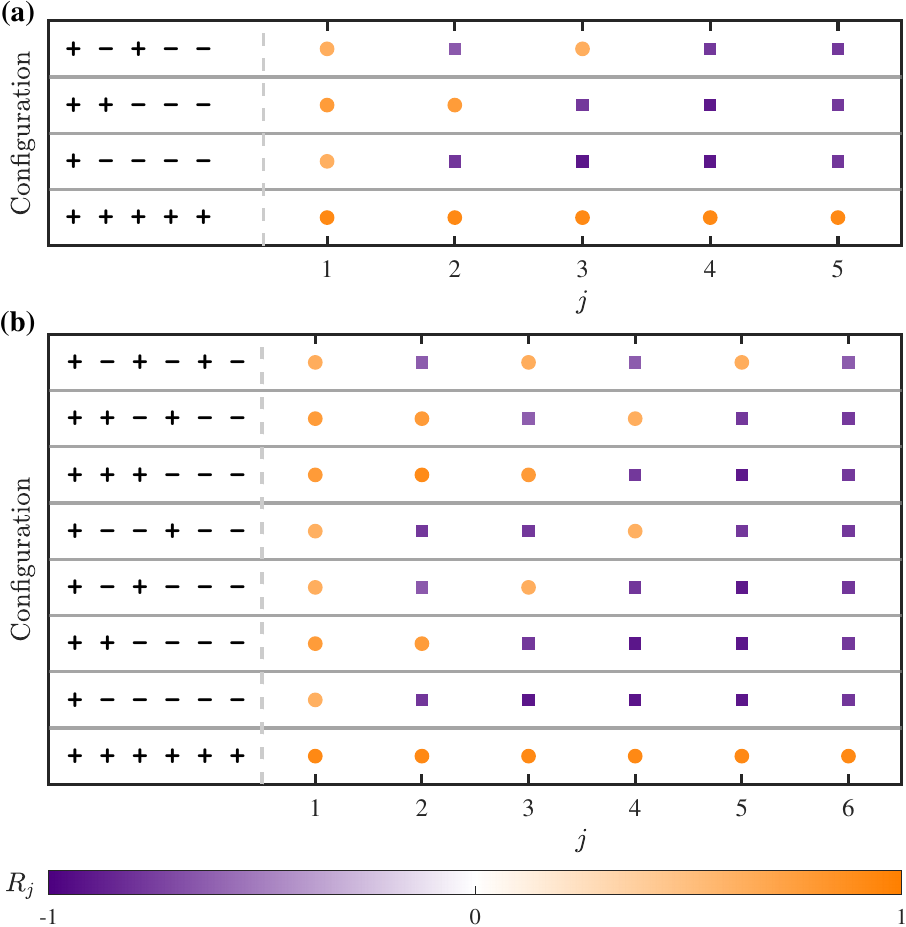}
    \caption{The representative configuration for (a) $N=5$ and (b) $N=6$. The parameters are set to $\omega_a=\omega_c=\omega$, $\kappa=0.4\omega$, $g=0.8\omega$ and $\xi=0.1\omega$.}
    \label{fig:N56}
\end{figure}

\begin{figure}
    \centering
    \includegraphics[width=\linewidth]{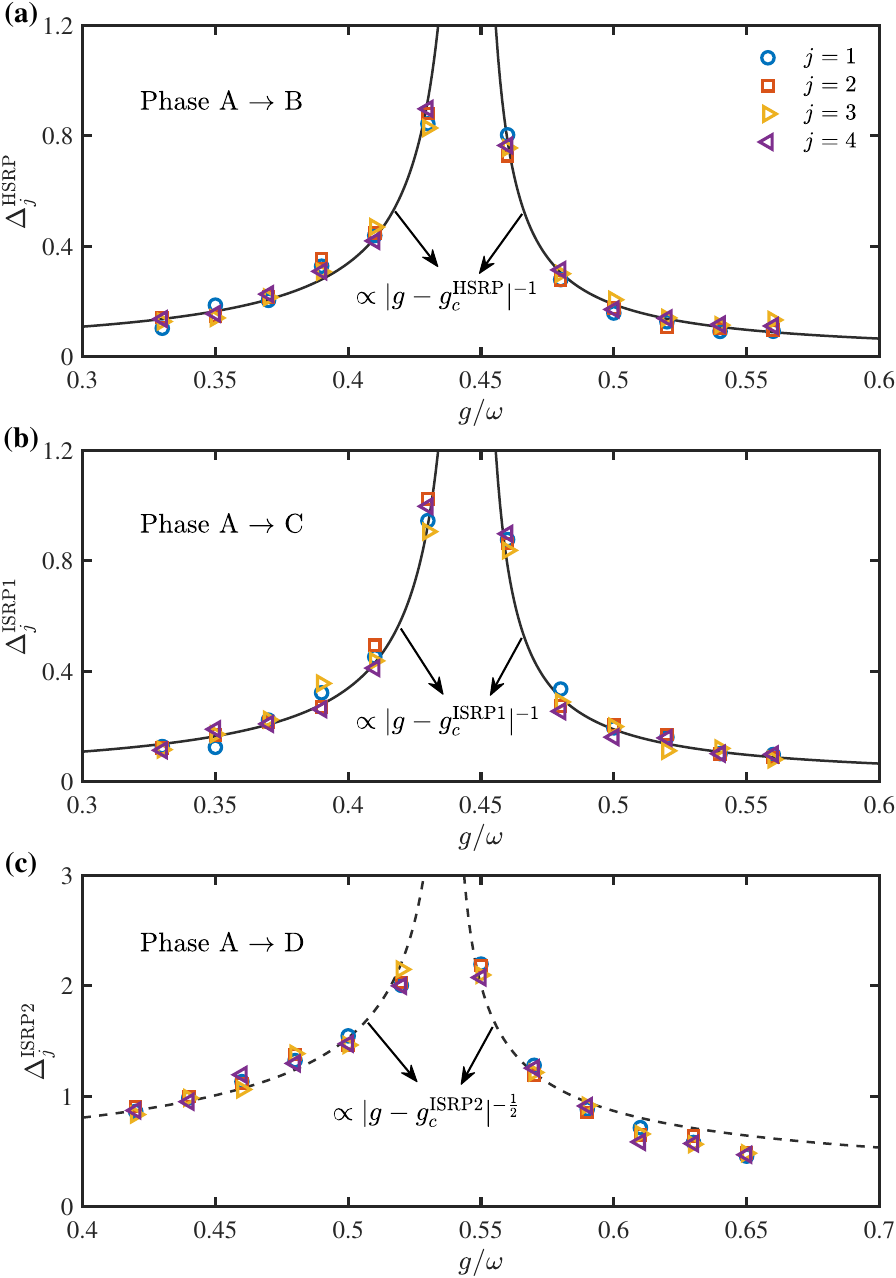}
    \caption{On-site fluctuations $\Delta_{j}$ in the vicinity of the critical coupling for (a) Phase A-B ($\xi=0.3\omega$), (b) Phase A-C ($\xi=-0.3\omega$), and (c) Phase A-D ($\xi=0.48\omega$). Other parameters are set to $\omega_a=\omega_c=\omega$ and $\kappa=0.4\omega$.}
    \label{fig:CS}
\end{figure}

The above analysis can be naturally generalized to larger Dicke lattices. For the cases of $N=5$ and $N=6$, the representative configuration classes, together with their degeneracies, are summarized in the last two rows of Table~\ref{tab:N}. Fig.~\ref{fig:N56} shows the corresponding representative steady-state configurations obtained in the superradiant regime. Compared with the four-site lattice, the number of distinct configuration classes increases significantly with the system size, leading to increasingly rich spatial ordering patterns of the SRP. Since different configuration classes may become dynamically stable simultaneously, larger Dicke lattices are expected to support even richer forms of multistability. Therefore, the configuration classification developed in this work provides a unified picture for understanding dissipative superradiant phases in Dicke lattices with arbitrary finite system sizes. Moreover, since the classification is based solely on the underlying lattice symmetry, the same strategy can also be applied to Dicke lattices with open boundary conditions after appropriately accounting for the corresponding spatial symmetries.

\subsection{Universality}
To characterize the universality of the dissipative phase transitions for the four-site Dicke lattice, we analyze the critical behavior of the on-site photon fluctuations~\cite{Nagy2011},
\begin{equation}
\Delta_j
=
\langle c_j^\dagger c_j\rangle
-
\langle c_j^\dagger\rangle\langle c_j\rangle.
\label{Delta}
\end{equation}
Within the mean-field approximation, the on-site fluctuations vanish identically and therefore cannot characterize the critical behavior. A full quantum treatment based on the Lindblad master equation can, in principle, capture these fluctuations, but rapidly becomes computationally prohibitive because the Hilbert-space dimension grows exponentially with both the number of lattice sites and the number of atoms. To overcome this difficulty, we employ the discrete truncated Wigner approximation (DTWA)~\cite{DTWA1,DTWA2,DTWA3,DTWA4}, which incorporates the leading quantum fluctuations through stochastic sampling of semiclassical trajectories.

In the DTWA, the resonator and atomic operators are mapped onto phase-space variables according to
$c_j\rightarrow\sqrt{N_a}(m_{R,j}^{c}+im_{I,j}^{c})$
and
$\sigma_{o,j}^{\mu}\rightarrow\sqrt{N_a}m_{o,j}^{\mu}$
($\mu=x,y,z$).
The corresponding stochastic equations of motion are
\begin{equation}
\begin{aligned}
d m_{R,j}^c
=&
\left[
-\kappa m_{R,j}^c
+\omega_c m_{I,j}^c
-\xi\left(m_{I,j+1}^c+m_{I,j-1}^c\right)
\right]dt\\
&+\sqrt{\frac{\kappa}{2N_a}}\,dW_{R,j}(t),\\
d m_{I,j}^c
=&
[-\omega_c m_{R,j}^c
-\kappa m_{I,j}^c
-\frac{g}{\sqrt{N_{a}}}\sum_{o=1}^{N_{a}} m^{x}_{o,j}\\
&+\xi\left(m_{R,j+1}^c+m_{R,j-1}^c\right)]dt+
\sqrt{\frac{\kappa}{2N_a}}\,dW_{I,j}(t),\\
d m^{x}_{o,j}
=&
-\omega_a m^{y}_{o,j}dt,\\
d m^{y}_{o,j}
=&
\left[
\omega_a m^{x}_{o,j}
-4g m_{R,j}^c m^{z}_{o,j}
\right]dt,\\
d m^{z}_{o,j}
=&
4g m_{R,j}^c m^{y}_{o,j}dt ,
\end{aligned}
\label{eq:DTWA}
\end{equation}
where $dW_{R,j}$ and $dW_{I,j}$ are independent real Wiener increments satisfying
$\langle dW_{\mu,j}\rangle=0$
and
$\langle dW_{\mu,j}dW_{\nu,l}\rangle
=
\delta_{\mu\nu}\delta_{jl}dt$
($\mu,\nu\in\{R,I\}$).
The stochastic noise originates from photon loss and represents the vacuum fluctuations associated with the Markovian reservoirs.

Physical observables are obtained by averaging over stochastic trajectories. In particular,
\begin{equation}
\begin{aligned}
\frac{\langle c_j\rangle}{\sqrt{N_a}}
&\simeq
\frac{1}{n_{t}}\sum_{i=1}^{n_{t}}(m_{R,i,j}^c+i m_{I,i,j}^c),\\
\frac{\langle c_j^{\dagger} c_{j}\rangle}{N_a}
&\simeq
\frac{1}{n_{t}}\sum_{i=1}^{n_{t}}[(m_{R,i,j}^c)^2+(m_{I,i,j}^c)^2]-\frac{1}{2N_a},
\end{aligned}
\end{equation}
from which the on-site photon fluctuation is evaluated according to Eq.~(\ref{Delta}). Here, $n_t$ denotes the number of stochastic trajectories. Throughout this work, we choose
$n_t\simeq1000$
and
$N_a=500$.

By solving the DTWA equations, we extract the critical behavior of the on-site photon fluctuations in the vicinity of the dissipative phase transitions. As shown in Fig.~\ref{fig:CS}, the fluctuations are well described by the scaling law
\begin{equation}
\Delta_j^\beta
\propto
|g-g_c^\beta|^{-\gamma_{\rm open}},
\end{equation}
where $g_c^\beta$ ($\beta=$ HSRP, ISRP1, ISRP2) denotes the corresponding critical coupling, and $\gamma_{\rm open}$ is the critical exponent characterizing the nonequilibrium universality class.

The extracted critical exponents reveal two distinct universality classes. As shown in Figs.~\ref{fig:CS}(a) and \ref{fig:CS}(b), the transitions from the NP to the HSRP and ISRP1 exhibit the same critical exponent,
$\gamma_{\rm open}=-1$,
indicating that they belong to the same universality class as the single dissipative Dicke model~\cite{Nagy2011}. In contrast, the transition from the NP to the ISRP2 is characterized by a different critical exponent,
$\gamma_{\rm open}=-1/2$,
as shown in Fig.~\ref{fig:CS}(c). This scaling agrees with that previously reported for the dissipative Dicke lattice with $N=3$ in the large-photon-hopping regime for open boundary conditions~\cite{NDickeWei}.

\section{Equilibrium phase transitions and universality}\label{closed}

In this section, we turn to the equilibrium phase transition of the closed Dicke lattice by setting $\kappa=0$. While the ground-state properties of Dicke lattices with an odd number of sites have recently been investigated~\cite{threeD2022}, the corresponding even-site lattices remain unexplored. Here, following the approach developed in Ref.~\cite{threeD2022}, we determine the ground-state phase diagram for the four-site Dicke lattice and analyze the associated superradiant phase transitions.

In the thermodynamic limit ($N_a\rightarrow\infty$), the Hamiltonian in Eq.~(\ref{eq:H}) can be reduced, within the mean-field approximation, to the form $\tilde{H}=E+H_q$,
where $E$ is the mean-field energy functional and $H_q$ is a quadratic Hamiltonian describing quantum fluctuations around the mean-field solution. The detailed derivation is presented in Appendix~\ref{Bclosed}. The mean-field energy is given by
\begin{equation}
\frac{E}{N_a} =
\sum_{j=1}^{N}
\left(
\omega_c\alpha_j^2
-
\frac{1}{2}
\sqrt{\omega_a^2+16g^2\alpha_j^2}
-
2\xi\alpha_j\alpha_{j+1}
\right),
\label{eq:EGS}
\end{equation}
where $\alpha_j=\langle c_j\rangle/\sqrt{N_a}$ is chosen to be real. By minimizing the energy functional with respect to
$\boldsymbol{\alpha}=(\alpha_1,\alpha_2,\cdots,\alpha_N)$,
we obtain both the ground-state energy $E_{\rm GS}$ and the corresponding superradiant configuration.

\begin{figure}
    \centering
    \includegraphics[width=\linewidth]{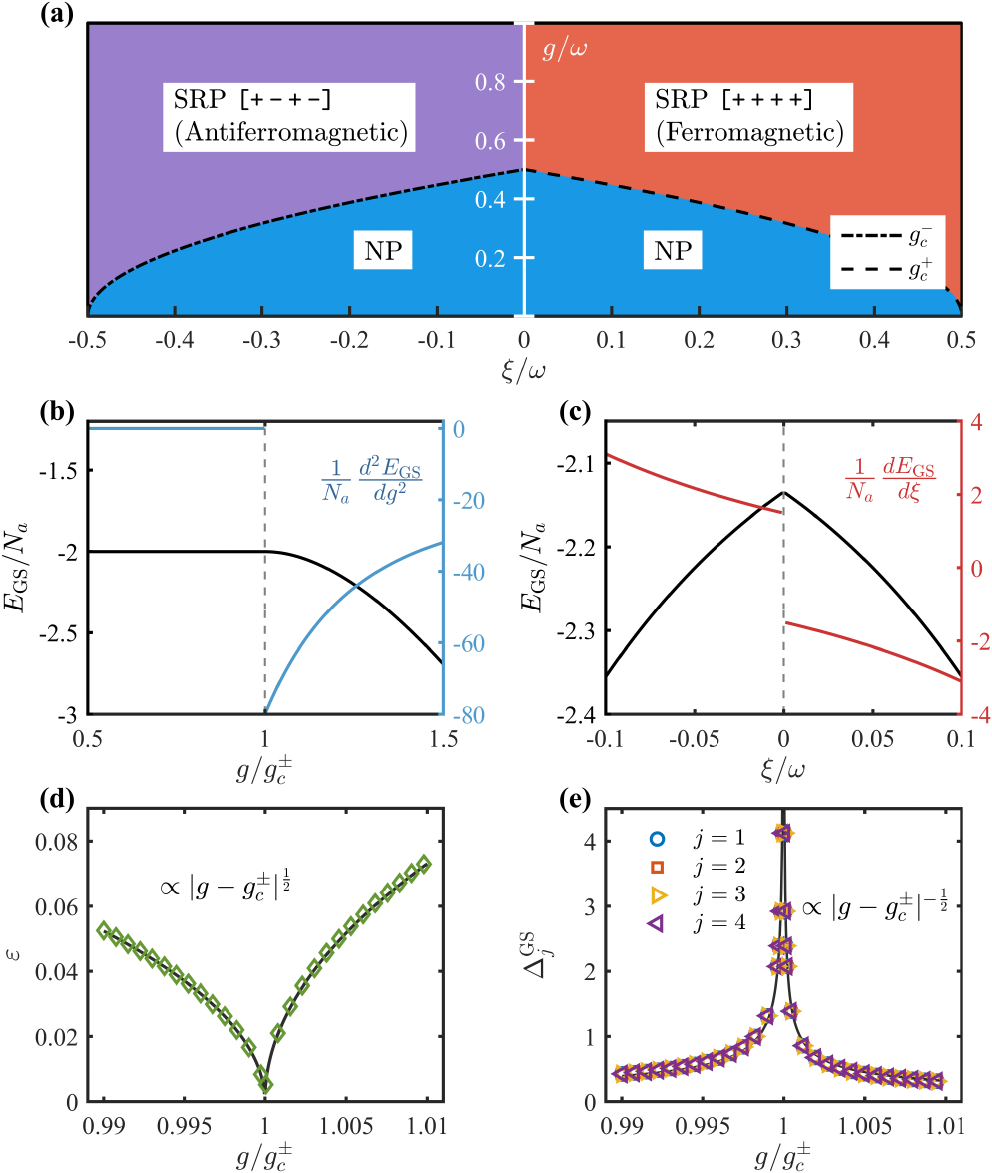}
    \caption{(a) Phase diagram for the Dicke lattice model with $N=4$. (b) The ground-state energy and its second-order derivative as a function of $g/g_{c}^{\pm}$. (c) The ground-state energy and its first-order derivative (red line) as a function of $\xi$ with $g=0.6\omega$. (d) The lowest excitation energies $\varepsilon$ as a function of the coupling strength $g/g_{c}^{\pm}$. (e) The on-site fluctuations of the ground state as a function of $g/g_{c}^{\pm}$. Other parameters are set to $\omega_a=\omega_c=\omega$ and $\xi=\pm 0.3\omega$.}
    \label{fig:closed}
\end{figure}

Following the derivation presented in Appendix~\ref{Bclosed} (see also Ref.~\cite{threeD2022}), we obtain the ground-state phase diagram for the four-site Dicke lattice, as shown in Fig.~\ref{fig:closed}(a). For positive (negative) photon hopping, the system undergoes a normal-to-superradiant phase transition when the atom--photon coupling exceeds the critical value
\begin{equation}
g_c^{\pm}
=
\frac{1}{2}
\sqrt{\omega_a(\omega_c\mp2\xi)}.
\end{equation}
For $\xi>0$, the superradiant phase is characterized by the homogeneous configuration
$\texttt{\textbf{[+\,+\,+\,+]}}$, corresponding to a ferromagnetic ordering pattern. In contrast, for $\xi<0$, the system enters the configuration
$\texttt{\textbf{[+\,-\,+\,-]}}$,
which corresponds to an antiferromagnetic ordering pattern. Therefore, unlike the dissipative Dicke lattice where different superradiant configurations may coexist, the equilibrium Dicke lattice always selects a unique ground-state configuration determined by the sign of the photon hopping. These two configurations correspond precisely to the representative classes listed in Table~\ref{tab:N}. Compared with the odd-site Dicke lattice~\cite{threeD2022}, where the negative-hopping regime supports a frustrated superradiant phase, the even-site lattice naturally favors the antiferromagnetic configuration, highlighting the intrinsic difference between odd and even lattice geometries.

Although the SRPs for $\xi>0$ and $\xi<0$ possess completely different spatial configurations, they share the same ground-state energy for identical values of $|\xi|$. This property is illustrated in Figs.~\ref{fig:closed}(b) and \ref{fig:closed}(c), where we plot the ground-state energy together with its derivatives. The results clearly show a continuous (second-order) phase transition from the NP to the SRP across $g_c^\pm$, whereas the transition between the ferromagnetic and antiferromagnetic SRPs across $\xi=0$ is of first order.

Finally, we characterize the universality of the equilibrium phase transition through the lowest excitation energy
$\epsilon$, defined as the smallest eigenvalue of the quadratic Hamiltonian $H_q$ (see Appendix~\ref{Bclosed}). As shown in Fig.~\ref{fig:closed}(d), the excitation gap closes according to the scaling law
$\epsilon\propto|g-g_c^\pm|^{\gamma_{\rm closed}}$
with the critical exponent
$\gamma_{\rm closed}=1/2$,
which coincides with the universality class of the single-site Dicke model~\cite{Dicke2003PRE}. The same universality is independently confirmed by the ground-state photon fluctuations shown in Fig.~\ref{fig:closed}(e), which satisfy
$\Delta_j^{\rm GS}\propto|g-g_c^\pm|^{-\gamma_{\rm closed}}$.
These results reveal a clear contrast between the equilibrium and nonequilibrium Dicke lattices: while different superradiant configurations lead to either same or distinct nonequilibrium universality classes in the dissipative system, they belong to the same equilibrium universality class in the closed Dicke lattice.

\section{Conclusion}\label{Con}

In summary, we have developed a configuration-based understanding of superradiant phase transitions in Dicke lattices. We show that the introduction of photon hopping naturally classifies the possible superradiant configurations according to the lattice symmetry, providing a unified perspective on the superradiant phases in both dissipative and closed Dicke lattice models. Within this framework, we determine the complete phase diagram of the four-site dissipative Dicke lattice, identify the coexistence of up to four stable superradiant phases, and further generalize the configuration classification to lattices with five and six sites.

Based on the proposed classification, we demonstrate that multistability in dissipative Dicke lattices originates from the simultaneous stabilization of different configuration classes. In contrast, for the closed Dicke lattice, the ground state uniquely selects one of the allowed configurations through energy minimization, giving rise to either the ferromagnetic or antiferromagnetic superradiant phase depending on the sign of the photon hopping. These results establish a unified physical picture connecting the equilibrium and nonequilibrium phase diagrams through the underlying superradiant configurations.

Furthermore, we investigate the critical behavior associated with different superradiant configurations. For the dissipative Dicke lattice, different configurations are shown to belong to either same or distinct nonequilibrium universality classes, whereas the corresponding equilibrium phase transitions in the closed Dicke lattice share the same universality class. This comparison reveals the fundamentally different roles played by dissipation in determining the critical behavior of superradiant phase transitions.

The present work highlights that the spatial organization of the superradiant order parameter provides a natural language for characterizing collective phases in Dicke lattices, suggesting that configuration may serve as a unifying concept for understanding phase transitions in a broad class of coupled light--matter systems.

\section*{Acknowledgments}

We thank Prof. Peter Rabl for his warm discussion. This work was supported by National Natural Science Foundation of China (Grant No. 12375010).

\section*{Data availability}

The data that support the findings of
this Letter are not publicly and are available from the authors upon reasonable request.

\appendix

\section{The steady state for open lattice with $N=4$}\label{Aopen}

The steady-state solutions
$\{R_j^{\rm ss},I_j^{\rm ss},(R_j^s)^{\rm ss},(I_j^s)^{\rm ss},(S_j^z)^{\rm ss}\}$
are obtained by setting the left-hand side of Eq.~(\ref{eq:ss}) to zero. To determine their dynamical stability, we introduce small fluctuations around the steady-state solutions. For example, we write
\[
R_j=R_j^{\rm ss}+\delta R_j,
\]
and substitute the corresponding expressions into Eq.~(\ref{eq:ss}). Retaining only terms that are linear in the fluctuations yields

\begin{equation}
\begin{aligned}
\delta \dot R_j
&=
-\kappa \delta R_j
+\omega_c \delta I_j
-\xi(\delta I_{j+1}+\delta I_{j-1}),
\\
\delta \dot I_j
&=
-\omega_c \delta R_j
-\kappa \delta I_j
-2g\delta Rs_j
+\xi(\delta R_{j+1}+\delta R_{j-1}),
\\
\delta \dot{Rs}_j
&=
\omega_a\delta Is_j,
\\
\delta \dot{Is}_j
&=
-\frac{\omega_a}
{(2Sz_j^{\rm ss})^2}
\delta Rs_j
+
4gSz_j^{\rm ss}\delta R_j .
\end{aligned}
\label{eq:F}
\end{equation}

Here, higher-order fluctuation terms have been neglected. In deriving Eq.~(\ref{eq:F}), we have used

\begin{equation}
\begin{aligned}
\delta Sz_j
&=
-\frac{Rs_j}{Sz_j}\delta Rs_j
-
\frac{Is_j}{Sz_j}\delta Is_j,
\\
R_j^{\rm ss}
&=
\frac{\omega_aRs_j^{\rm ss}}
{4gSz_j^{\rm ss}},
\qquad
Is_j^{\rm ss}=0,
\end{aligned}
\end{equation}
which follow from the conservation of the spin length and the steady-state conditions of Eq.~(\ref{eq:ss}), respectively.

Equation~(\ref{eq:F}) can be written compactly as $\dot{\boldsymbol{\Psi}}=M\boldsymbol{\Psi}$,
where $\boldsymbol{\Psi}=(\psi_1,\psi_2,\psi_3,\psi_4)^T$, with $\psi_j=(\delta R_j,\delta I_j,\delta Rs_j,\delta Is_j)^T$, and the coefficient matrix is
\begin{equation}
M=
\begin{pmatrix}
M_{1} & M_{\xi} & 0 & M_{\xi}\\
M_{\xi} & M_{2} & M_{\xi} & 0\\
0 & M_{\xi} & M_{3} & M_{\xi}\\
M_{\xi} & 0 & M_{\xi} & M_{4}
\end{pmatrix}.
\label{eq:Mtot}
\end{equation}
The diagonal and off-diagonal $4\times4$ blocks are given by
\begin{equation}
M_j=
\begin{pmatrix}
-\kappa & \omega_c & 0 & 0\\
-\omega_c & -\kappa & -2g & 0\\
0 & 0 & 0 & \omega_a\\
4gSz_j^{\rm ss} &
0 &
-\dfrac{\omega_a}
{(2Sz_j^{\rm ss})^2}
&
0
\end{pmatrix},
\end{equation}
and
\begin{equation}
M_{\xi}
=
\begin{pmatrix}
0 & -\xi & 0 & 0\\
\xi & 0 & 0 & 0\\
0 & 0 & 0 & 0\\
0 & 0 & 0 & 0
\end{pmatrix},
\end{equation}
respectively.

According to the Routh--Hurwitz criterion~\cite{QZZL1987}, a steady-state solution is dynamically stable if and only if all eigenvalues of the coefficient matrix $M$ have negative real parts. For the analytical steady-state solutions discussed in the main text, namely the NP, HSRP, ISRP1, and ISRP2, the steady-state atomic population is identical at every lattice site. For example, the ISRP1 satisfies
\[
(Sz_1)^{\rm ss}
=
(Sz_2)^{\rm ss}
=
(Sz_3)^{\rm ss}
=
(Sz_4)^{\rm ss}
=
\mathcal{Z}_2.
\]
Consequently, the coefficient matrix $M$ possesses the translational symmetry of the lattice and can be block diagonalized by a discrete Fourier transformation,
\begin{equation}
U^{\dagger}MU
=
\operatorname{diag}
(M_1,M_2,M_3,M_4).
\end{equation}
Here,
\[
U=
[\nu_{1}\otimes\mathbb{I}_{4},
\nu_{2}\otimes\mathbb{I}_{4},
\nu_{3}\otimes\mathbb{I}_{4},
\nu_{4}\otimes\mathbb{I}_{4}],
\]
where $\mathbb{I}_{4}$ denotes the $4\times4$ identity matrix, and
$\{\nu_k\}$ is the discrete Fourier basis,
\begin{equation}
\nu_k
=
\frac{1}{2}
\left(
1,
e^{-i\frac{\pi}{2}(k-1)},
e^{-i\pi(k-1)},
e^{-i\frac{3\pi}{2}(k-1)}
\right)^T.
\label{eq:PBSU}
\end{equation}
The stability analysis is therefore reduced to four independent
$4\times4$ matrices,
\begin{equation}
M_k
=
M_j
+
2\cos
\!\left[
\frac{\pi(k-1)}{2}
\right]
M_{\xi},
\end{equation}
where $k=1,2,3,4$.

According to the Routh--Hurwitz criterion, the steady-state solution is dynamically stable provided that all eigenvalues of every $M_k$ have negative real parts, implying that $\operatorname{Det}(M_k)>0$ is a necessary condition. The corresponding determinants are
\begin{equation}
\begin{aligned}
\operatorname{Det}
(M_k^{\rm NP})
&=
\omega_a^2
(\kappa^2+\omega_k^2)
-
4g^2
\omega_a
\omega_k,
\\
\operatorname{Det}
(M_k^{\rm HSRP})
&=
\frac{
16g^4
\omega_1^2
(\kappa^2+\omega_k^2)
}{
(\kappa^2+\omega_1^2)^2
}
-
\frac{
\omega_a^2
\omega_k
(\kappa^2+\omega_1^2)
}{
\omega_1
},
\\
\operatorname{Det}
(M_k^{\rm ISRP1})
&=
\frac{
16g^4
\omega_3^2
(\kappa^2+\omega_k^2)
}{
(\kappa^2+\omega_3^2)^2
}
-
\frac{
\omega_a^2
\omega_k
(\kappa^2+\omega_3^2)
}{
\omega_3
},
\\
\operatorname{Det}
(M_k^{\rm ISRP2})
&=
\frac{
16g^4
\omega_c^2
(\kappa^2+\omega_k^2)
}{
(\kappa^2+\omega_c^2)^2
}
-
\frac{
\omega_a^2
\omega_k
(\kappa^2+\omega_c^2)
}{
\omega_c
},
\end{aligned}
\end{equation}
which yields the analytical expressions of the critical couplings for the NP, HSRP, ISRP1, and ISRP2.

\section{The ground state of the closed Dicke lattice}\label{Bclosed}

In this Appendix, we briefly outline the procedure for determining the ground-state properties of the closed Dicke lattice, following the approach developed in Ref.~\cite{threeD2022}.

The mean-field expectation values of the resonator field and the collective spin at site $j$ are parameterized as $\langle c_j\rangle=\sqrt{N_a}\alpha_j$,
and $(\langle S_j^x\rangle,\langle S_j^y\rangle,\langle S_j^z\rangle)
=N_a(\sin\theta_j\cos\phi_j,\sin\theta_j\sin\phi_j,\cos\theta_j)/2$
where $\theta_j\in[0,\pi]$ and $\phi_j\in[0,2\pi)$.

In the thermodynamic limit
($N_a\rightarrow\infty$),
the Hamiltonian in Eq.~(\ref{eq:H}) is transformed according to $\tilde H
=U^\dagger H U$, with
\begin{equation}
U
=
\prod_{n=1}^{N}
e^{-i\phi_nS_n^z}
e^{-i\theta_nS_n^y}
e^{\sqrt{N_a}
(\alpha_nc_n^\dagger-\alpha_nc_n)}.
\end{equation}
The Holstein--Primakoff transformation is then applied to the rotated collective spins,
\[
S_j^+
=
\sqrt{N_a-b_j^\dagger b_j}\,b_j,
\qquad
S_j^z
=
\frac{N_a}{2}
-
b_j^\dagger b_j,
\]
where the bosonic operators satisfy
$[b_j,b_j^\dagger]=1$.

Consequently,
$\tilde H$
contains a constant term together with terms that are linear, quadratic, and higher order in the bosonic operators
$c_j$
and
$b_j$.
The higher-order terms vanish in the thermodynamic limit and are therefore neglected. Choosing $\alpha_j\in\mathbb{R}$ and
\begin{equation}
\begin{aligned}
\cos\phi_j
&=
-\alpha_j/|\alpha_j|,
\qquad
\sin\phi_j=0,
\\
\cos\theta_j
&=
-\frac{\omega_a}
{\sqrt{\omega_a^2+16g^2\alpha_j^2}},
\\
\sin\theta_j
&=
\frac{4g|\alpha_j|}
{\sqrt{\omega_a^2+16g^2\alpha_j^2}},
\end{aligned}
\end{equation}
eliminates all linear terms in the transformed Hamiltonian.

The ground-state energy
$E_{\rm GS}$
is therefore obtained by minimizing Eq.~(\ref{eq:EGS}) in the main text with respect to $\boldsymbol{\alpha}$. In the NP, $\boldsymbol{\alpha}=0$, and the ground-state energy is $E_{\rm GS}/N_a=-N\omega_a/2$. To determine the stability of the NP, we evaluate the Hessian matrix of $E/N_a$ at $\boldsymbol{\alpha}=0$,
\begin{equation}
{\rm Hess}_{jk}
=
\left.
\frac{\partial^2(E/N_a)}
{\partial\alpha_j\partial\alpha_k}
\right|_{\boldsymbol{\alpha}=0}.
\end{equation}
Its eigenvalues are
\[
\lambda_k
=
2\omega_c
-
\frac{8g^2}{\omega_a}
-
4\xi
\cos
\left(
\frac{2\pi k}{N}
\right),
\]
where
$k=0,1,\cdots,N-1$.

The NP remains stable only when all eigenvalues satisfy
$\lambda_k>0$.
For the four-site lattice, this condition gives the critical coupling strengths
\begin{equation}
g_c^\pm
=
\frac12
\sqrt{\omega_a(\omega_c\mp2\xi)}.
\end{equation}
Therefore, the NP becomes unstable when
$g>g_c^+$
for
$\xi>0$
or
$g>g_c^-$
for
$\xi<0$,
signaling the onset of the SRP.

In the SRP, the ground-state energy has the same analytical form for both positive and negative photon hopping,
\begin{equation}
\frac{E_{\rm GS}^{+}}{N_a}
=
-\frac{N}{2}
\left[
\frac{2g^2}
{\omega_c-2|\xi|}
+
\frac{\omega_a^2(\omega_c-2|\xi|)}
{8g^2}
\right].
\label{eq:EGS+}
\end{equation}
The corresponding superradiant configurations are $\alpha_1
=\alpha_2=\alpha_3=\alpha_4=\alpha$ for $\xi>0$, and $\alpha_1=-\alpha_2=\alpha_3=-\alpha_4=\alpha$ for $\xi<0$, where
\begin{equation}
\alpha
=
\pm
\frac12
\sqrt{
\frac{4g^2}
{\omega_c-2|\xi|}
-
\frac{\omega_a^2}
{4g^2}
}.
\end{equation}

The quadratic Hamiltonian describing the elementary excitations is
\begin{equation}
\begin{aligned}
H_q
=
&
\sum_{j=1}^{N}
\Big[
\omega_cc_j^\dagger c_j
-
\xi
(c_j^\dagger c_{j+1}
+
c_jc_{j+1}^\dagger)
\\
&
-
\frac{\omega_a}{\cos\theta_j}
b_j^\dagger b_j
+
g\cos\phi_j\cos\theta_j
(c_j^\dagger+c_j)
(b_j^\dagger+b_j)
\Big].
\end{aligned}
\end{equation}
Introducing the resonator quadratures
$q_j=(c_j+c_j^\dagger)/\sqrt2$,
$p_j=i(c_j^\dagger-c_j)/\sqrt2$,
and the atomic quadratures
$Q_j=(b_j+b_j^\dagger)/\sqrt2$,
$P_j=i(b_j^\dagger-b_j)/\sqrt2$,
the quadratic Hamiltonian can be brought into a diagonal form through a generalized Bogoliubov transformation. The detailed diagonalization procedure is identical to that presented in Ref.~\cite{threeD2022} and is therefore not repeated here. The excitation spectrum and the ground-state photon fluctuations shown in the main text are obtained directly from the resulting normal modes.


\end{document}